\documentclass[twocolumn,aps,prx]{revtex4-1}

\usepackage{graphics}
\usepackage{graphicx}
\usepackage{dcolumn}
\usepackage{bm}
\usepackage{amsmath,amssymb,epstopdf,color}

\begin{document}


\title{The Hawking Temperature of Anti-de Sitter Black Holes: Topology and Phase Transitions}
\author{Charles W. Robson, Leone Di Mauro Villari and Fabio Biancalana}
\affiliation{School of Engineering and Physical Sciences, Heriot-Watt University, EH14 4AS Edinburgh, UK}
\date{\today}

\begin{abstract}

In this work we determine how the description of a four-dimensional Schwarzschild-anti de Sitter black hole affects the topological calculation of its Hawking temperature. It is shown that a two-dimensional approach is required due to the presence of the Hawking-Page phase transition which destabilises the spacetime's topology. We prove that a dimensional reduction removes the phase transition and hence stabilises the system. This hints at a previously unknown feature of black hole thermodynamics, namely that certain black holes may demand a lower-dimensional description in order to define their Hawking temperatures.

\end{abstract}

\maketitle

\section{Introduction}

It has been recently shown by the authors of this paper \cite{ourpaper1} that the Hawking temperatures \cite{Hawking_rad} of black holes can be found using purely topological methods. This has led to a deeper understanding of the topological nature of the Hawking temperature of not only individual black hole event horizons, but also multihorizon systems, as well as cosmological horizons such as that found in pure de Sitter spacetime \cite{ourpaper2,ovgun}.

The foundation of the method rests on a familiar topological invariant known as the Euler characteristic $\chi$, which can be defined for Euclideanised black hole spacetimes \cite{Chern1,Chern2,Gibbons}. Using this invariant, plus some extra basic information about black hole spacetimes, such as their curvature scalars and Killing horizon structure, allows for Hawking temperatures to be easily calculated. This has been shown to be reliable for a wide variety of black hole systems in both two and four dimensions and is independent of coordinate system choice \cite{ourpaper1,ourpaper2}.

For our four-dimensional world, the most physically-relevant targets of study are four-dimensional black holes. It has recently been shown that both two- and four-dimensional topological approaches can be used to study a large collection of four-dimensional black holes, in every case investigated until now giving completely equivalent results \cite{ourpaper1,ourpaper2}. In cases where a four-dimensional topological study is impeded by technical problems (for example by a vanishing Euler characteristic), a dimensional reduction to two dimensions works astonishingly well, even for black holes with highly nontrivial topologies, suggesting that two-dimensional descriptions may be sufficient for describing the thermodynamical properties of all black holes. This thermodynamical information storage in two dimensions is reminiscent of the well-known area scaling law of black hole entropy, as well as more speculative research on the holographic nature of the Universe as a whole \cite{Bekenstein,Susskind,Susskind2}.

The present work demonstrates that due to the Hawking-Page phase transition for a black hole in four-dimensional anti-de Sitter ($\rm{AdS_{4}}$) spacetime, two- and four-dimensional topological approaches give conflicting Hawking temperature results, signifying that these black holes constitute a special case deserving of careful study. We show that a topological analysis of the Hawking temperature of a Schwarzschild-AdS (S-$\rm{AdS_{4}}$) black hole {\em demands} a two-dimensional approach, despite the S-$\rm{AdS}_{4}$ black hole being a four-dimensional object. Our analysis shows that this deviation is caused by the Hawking-Page phase transition, manifesting as a topological instability.

This instability is, however, a blessing in disguise, as it sheds more light on the intrinsic two-dimensional character of black holes. A simple dimensional reduction of the S-$\rm{AdS_{4}}$ metric proves sufficient in yielding the correct Hawking temperature. The reduction succeeds because it removes all information concerning the Hawking-Page phase transition, in effect stabilising the system's topology.

The plan of the paper is as follows. In Section \ref{sec:Sec1} we reintroduce our two- and four-dimensional topological methods for calculating Hawking temperatures. These two approaches have in the past given equivalent results for black hole spacetimes \cite{ourpaper1,ourpaper2}, however we show in Section \ref{sec:Sec2} that this is not the case in an anti-de Sitter background. In Section \ref{sec:Sec3} we prove using a minimisation procedure that the reason for this discrepancy is the presence of the Hawking-Page phase transition, which introduces a topological instability. Reducing the description of the S-$\rm{AdS}_{4}$ black hole down to two dimensions is shown to eliminate any information regarding the phase transition and hence stabilise the system, accounting for the correct Hawking temperature yielded by the two-dimensional approach. Finally, an Appendix treats other members of the AdS family of black holes.

\section{Calculating the Hawking temperature of  black holes in two and four dimensions} \label{sec:Sec1}

We reproduce here the formula for the Hawking temperature of a two-dimensional (or four-dimensional, after reduction) black hole with a time-independent Ricci scalar $R$:
\begin{equation} \label{eq:2d_form}
T_{\mathrm{H}}=\frac{\hbar c}{4\pi \chi k_{\mathrm{B}}}\sum_{j \leq \chi}\int_{r_{\mathrm{H_{j}}}}\sqrt{g}Rdr,
\end{equation}
where $k_{\rm B}$ is Boltzmann's constant, $c$ is the speed of light in vacuum, $\hbar$ the reduced Planck constant, $R(r)$ the Ricci scalar depending only on the `spatial' variable $r$, $r_{\rm H_{j}}$ the location of the $j$-th Killing horizon, $g$ the (Euclidean) metric determinant, and $\chi$ the Euler characteristic of the black hole's Euclideanised spacetime. The symbol $\sum_{j \leq \chi}$ is a sum over all Killing horizons, where one should pay attention to the sign of each term in the sum, as it can be positive or negative depending on the specific features of each horizon -- the overall temperature result is of course always positive \cite{ourpaper1}.

A four-dimensional version of the above Hawking temperature expression cannot be found in such simple and closed form. This is because the value of $\chi$ in four dimensions is not simply equal to the number of fixed points of a Killing vector field on the manifold (as holds in two dimensions) leading to complications on how to sum correctly over horizons in any potential temperature expression \cite{Hawking}. A closed expression for a four-dimensional black hole temperature would therefore be much more complicated, if indeed possible. This lack of a general expression is not a problem for practical calculations. The Euler characteristic of a four-dimensional manifold is
\begin{equation} \label{eq:4d_Euler}
\chi=\frac{1}{32\pi^2}\int d^4 x \sqrt{g}\left( K_{1} - 4R_{ab}R^{ab} + R^2 \right),
\end{equation}
where $R_{ab}$ is the Ricci tensor, $K_{1}$ the Kretschmann invariant, and $g$ the Euclidean metric determinant. Using the above expression, a Hawking temperature formula on an ad hoc basis for each four-dimensional spacetime, can be directly found \cite{ourpaper2}.

A detailed treatment of this last point, along with a full derivation of formula (\ref{eq:2d_form}) and the necessary topological background, can be found in Refs. \cite{ourpaper1,ourpaper2,Chern1,Chern2,Gibbons,Hawking,Eguchi,Morgan}.

\section{Two- and four-dimensional topology of black holes} \label{sec:Sec2}

In this work, we show for the first time that the Hawking temperature of a black hole in $\rm{AdS_{4}}$ spacetime cannot be found using a four-dimensional topological approach due to a topological instability seeded by the Hawking-Page phase transition.

This phase transition causes S-$\rm{AdS_{4}}$ black holes to be thermodynamically unfavoured below a certain critical temperature \cite{Page,Ong}. We show that a very simple dimensional reduction down to two dimensions suppresses the phase transition. Therefore, any information pertaining to the thermodynamical, and therefore topological, instability of the system is removed after this reduction, allowing for a two-dimensional topological calculation of the black hole's Hawking temperature using formula (\ref{eq:2d_form}).

In this work we also show that dimensional reduction allows topological Hawking temperature calculations to be performed for other, more exotic, members of the $\rm{AdS_{4}}$ black hole family, including those with toral and hyperbolic horizons. These results are presented in the Appendix.

As a first step, let us directly calculate an S-$\rm{AdS_{4}}$ black hole's Hawking temperature using formula (\ref{eq:2d_form}). The Euclideanised S-$\rm{AdS_{4}}$ metric is given by \cite{Charm}
\begin{widetext}
\begin{equation} \label{S-AdS4}
ds^2 = \left( 1+\frac{r^2}{L^2} - \frac{2M}{r} \right) dt^2 + \left( 1+\frac{r^2}{L^2} - \frac{2M}{r} \right)^{-1} dr^2 + r^2 d\Omega^2,
\end{equation}
\end{widetext}
where $L$ is the AdS curvature length scale, related (in four dimensions) to the cosmological constant by $\Lambda=-3/L^2$; $M$ is the hole's mass.

As its name suggests, the metric for the S-$\rm{AdS_{4}}$ black hole reverts to that of the Schwarzschild black hole in the $\Lambda \rightarrow 0$ limit. The S-$\rm{AdS_{4}}$ black holes we study in the main text have spherical horizons and in the Appendix we analyse their flat and hyperbolic cousins.

Dimensionally reducing metric (\ref{S-AdS4}) in the crudest way possible, by excising the two-sphere section from the line element, remarkably lets one use formula (\ref{eq:2d_form}) yielding the correct Hawking temperature result \cite{Ong}:
\begin{equation} \label{correct_temp}
T_{\rm{H}}=\frac{L^2 + 3r_{h}^2}{4\pi r_{h}L^2},
\end{equation}
where $r_{h}$ is the AdS black hole radius.

Using a four-dimensional topological formula, which can be derived directly from the Euler characteristic expression (\ref{eq:4d_Euler}), produces a Hawking temperature result which differs from the correct temperature (\ref{correct_temp}) \cite{ourpaper2}. From metric (\ref{S-AdS4}), the geometric quantities required to evaluate $\chi$ from (\ref{eq:4d_Euler}) can be immediately found, i.e. $\sqrt{g}=r^2 \rm{sin}(\theta)$ and $K_{1} - 4R_{ab}R^{ab} + R^2=24\left( 1/L^4 + 2M^2/r^6 \right)$. Substituting these values into (\ref{eq:4d_Euler}) and integrating over the two-sphere gives \cite{Gibbons} $\chi=(3/\pi)\int_{0}^{\beta_{2}} dt \int_{r_{h}} dr \left( r^2/L^4 + 2M^2/r^4 \right)$, where $\beta_{2}$ is the inverse of the black hole's Hawking temperature. One can then integrate over time and space leading to $\chi T_{\rm{H}} = \left( -r_{h}^6/L^4 + 2M^2 \right)/(r_{h}^3 \pi)$. Metric (\ref{S-AdS4}) has time isometry, with associated Killing vector, and has one real-valued fixed point two-surface at $r_{h}$; this single ``bolt" gives a value for this space of $\chi=2$ \cite{Hawking}. At the event horizon, the $g_{tt}$ component of (\ref{S-AdS4}) vanishes, providing an expression for the black hole's mass: $M=r_{h}\left( L^2 + r_{h}^2 \right)/(2L^2)$. Substituting the mass and $\chi$ values into our previous expression for $\chi T_{\rm{H}}$ finally gives
\begin{equation} \label{eq:wrong1}
T_{\rm{H}}=\frac{L^2 + 2r_{h}^2 - r_{h}^4/L^2}{4\pi r_{h}L^2};
\end{equation}
this is different from the correct value (\ref{correct_temp}).

The reason for this discrepancy is that S-$\rm{AdS_{4}}$ black holes are known to suffer from a thermodynamic instability via the Hawking-Page phase transition \cite{Page}. This phase transition leads to a violent topology change in the spacetime manifold, with the Euler characteristic's value transitioning from non-zero to zero below a critical temperature. The vanishing of $\chi$ is something our topological methods are unable to deal with due to $\chi$'s presence in the denominator of all temperature formulas.

The Hawking-Page phase transition demonstrates that S-$\rm{AdS_{4}}$ black holes below a critical temperature become thermodynamically unfavoured, with radiation in thermal equilibrium in $\rm{AdS_{4}}$ space favoured \cite{Page,Ong}. Crucially, Euclideanised $\rm{AdS_{4}}$ space containing only radiation has an Euler characteristic of $\chi=0$, whereas Euclideanised S-$\rm{AdS_{4}}$ space has $\chi=2$; for further details on each Euclidean space's topology, see \cite{Witten}. Therefore, below the critical temperature, a manifold with $\chi=0$ is thermodynamically favoured over that with $\chi=2$. Whenever $\chi=0$ one would expect our finite $\chi$-based temperature method to break down and, indeed, this is the case.

Let us now plot the temperature form (\ref{eq:wrong1}) along with the correct, known result (\ref{correct_temp}) to see how they differ, i.e. to pinpoint where formula (\ref{eq:wrong1}) breaks down and where the two- and four-dimensional approaches diverge.
\begin{figure}
\centering
\includegraphics[width=80mm]{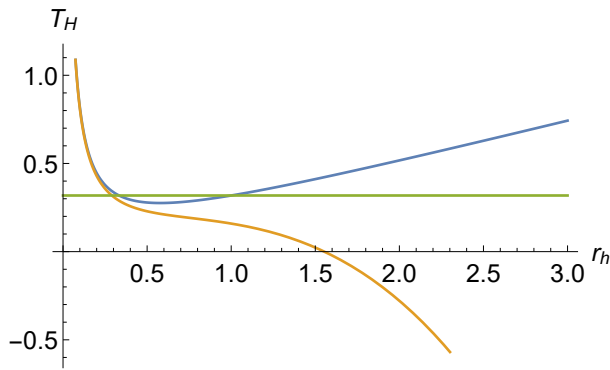}
\caption{The orange line shows the incorrect S-$\rm{AdS_{4}}$ black hole Hawking temperature as a function of horizon radius predicted by (\ref{eq:wrong1}). The blue line is the correct, known form verified by (\ref{correct_temp}). The green line is at $T_{crit}$, the critical temperature below which S-$\rm{AdS_{4}}$ black holes become unfavourable thermodynamically. The AdS scale $L$ has been set to one for simplicity and all units are arbitrary.}
\label{fig:wrong1}
\end{figure}
As can be clearly seen in Fig. \ref{fig:wrong1}, (\ref{correct_temp}) and (\ref{eq:wrong1}) match very closely at temperatures above the critical temperature (marked by the green line) on the small $r_{h}$ branch. At the critical temperature the Hawking-Page phase transition occurs and below it the spacetime tends to thermal AdS with $\chi=0$, a region in which our formulas are no longer applicable, evinced by the diverging plots.

One may ask: why does the two-dimensional formula (\ref{eq:2d_form}) give the correct S-$\rm{AdS_{4}}$ Hawking temperature despite the presence of $\chi$ in its denominator? Surely $\chi$ will vanish at temperatures below the Hawking-Page phase transition, creating a divergence? The reason (\ref{eq:2d_form}) functions is that a dimensional reduction down to two dimensions removes all information connected with the phase transition, stabilising the thermodynamics and hence the topology as always having a finite $\chi$ value. In the next section we prove this.

\section{The Hawking-Page phase transition} \label{sec:Sec3}

The free energy of a Euclideanised spacetime is known to be \cite{Ong}
\begin{equation} \label{free_energy}
F=-T\mathrm{log}Z \approx TS,
\end{equation}
where $Z$ is the partition function, and $TS$ the product of temperature and the Euclidean action. Using (\ref{free_energy}) one can see that for two Euclideanised spacetimes {\em at the same temperature} the one with the lower Euclidean action will have a lower free energy and hence will be thermodynamically favoured.

We now derive the Hawking-Page phase transition in four dimensions by comparing the Euclidean actions of two spacetimes: $\rm{AdS_{4}}$ space with and without a black hole \cite{Page,Ong}. This standard method will then be applied to a dimensionally-reduced spacetime, explaining the efficacy of our topological temperature formula (\ref{eq:2d_form}).

The Einstein-Hilbert action has both Lorentzian and Euclidean versions, differing from each other by a global minus sign. The four-dimensional Euclidean action is given by
\begin{equation} \label{Eucl_act}
S_{E}=-\frac{1}{16\pi}\int d^4 x \sqrt{-g}\left( R-2\Lambda \right).
\end{equation}
Using (\ref{Eucl_act}), the Euclidean actions of S-$\rm{AdS_{4}}$ and $\rm{AdS_{4}}$ spacetimes can be compared, identifying the Hawking-Page phase transition which occurs when these two actions become equal \cite{Page}.

The Euclidean actions can be derived straight from Lorentzian metrics. Firstly, $\rm{AdS_{4}}$ spacetime can be described by \cite{Ong}
\begin{equation}
ds^2 = -\left( 1+\frac{r^2}{L^2} \right) dt^2 + \left( 1+\frac{r^2}{L^2} \right)^{-1} dr^2 + r^2 d\Omega^2,
\end{equation}
which, using (\ref{Eucl_act}), gives a Euclidean action of
\begin{equation}
S_{\mathrm{EAdS}}=-\frac{\Lambda}{8\pi}\int d^4 x \sqrt{-g} = -\frac{\Lambda}{6}\beta_{1}\xi^3,
\end{equation}
where the radial integral $r$ was evaluated from zero to a cutoff $\xi$ to ensure finiteness, and time $t$ was integrated from zero to an arbitrary period $\beta_{1}$, the value of which can later be fixed by an asymptotic condition.

The S-$\rm{AdS_{4}}$ black hole spacetime, described by
\begin{widetext}
\begin{equation} \label{SAdS_Lorentz}
ds^2 = -\left( 1+\frac{r^2}{L^2} - \frac{2M}{r} \right) dt^2 + \left( 1+\frac{r^2}{L^2} - \frac{2M}{r} \right)^{-1} dr^2 + r^2 d\Omega^2,
\end{equation}
\end{widetext}
has Euclidean action
\begin{equation}
S_{\mathrm{EAdS-Sch}} = -\frac{\Lambda}{6}\beta_{2}\left( \xi^3 - r_{h}^3 \right),
\end{equation}
where the radial integral was taken between $r_{h}$ and the cutoff $\xi$, and over time period $\beta_{2}$ (the inverse of the S-$\rm{AdS_{4}}$ black hole's temperature).

The Hawking-Page phase transition can now be derived by finding where the two actions defined above become equal, at fixed temperature. The metric tensors must match at the cutoff $\xi$, whose limit will be taken as $\xi \rightarrow \infty$, enforcing the same asymptotic geometry for the two metrics \cite{Page,Ong}. This asymptotic constraint fixes the time period $\beta_{1}$ by
\begin{equation}
\beta_{1}\sqrt{1+\frac{\xi^2}{L^2}}=\beta_{2}\sqrt{1-\frac{2M}{\xi}+\frac{\xi^2}{L^2}}.
\end{equation}
The difference between the two actions at the asymptotic limit is
\begin{equation}
\lim_{\xi \rightarrow \infty} \left( S_{\mathrm{EAdS-Sch}} - S_{\mathrm{EAdS}} \right) = -\frac{\beta_{2}\left( r_{h}^3 - ML^2 \right)}{2L^2}.
\end{equation}
Clearly, when this difference in actions is positive then thermal $\rm{AdS_{4}}$ space becomes thermodynamically favoured over the existence of a black hole. This occurs when $r_{h}^3 < ML^2$. Substituting in the S-$\rm{AdS_{4}}$ black hole mass defined earlier leads to the constraint $r_{h}^2 < L^2$. It is at the radius $r_{h}=L$ that the black hole's temperature equals $T_{\mathrm{H}}=T_{crit}=1/(\pi L)$; this is where the Hawking-Page phase transition occurs. Interestingly, unlike the Schwarzschild black hole, the S-$\rm{AdS_{4}}$ black hole has a minimum temperature $T_{min}=\sqrt{3}/(2\pi L)$; it is between temperatures $T_{crit}$ and $T_{min}$ that pure radiation in $\rm{AdS_{4}}$ space is thermodynamically favoured over the existence of a black hole.

We now adapt the above proof to a dimensionally-reduced two-dimensional action showing that the phase transition vanishes. This explains the predictive power of equation (\ref{eq:2d_form}). The S-$\rm{AdS_{4}}$ metric is dimensionally reduced by simply removing the angular section of metric (\ref{SAdS_Lorentz}) to obtain
\begin{equation} \label{eq:trunc}
ds^2 = -\left( 1+\frac{r^2}{L^2} - \frac{2M}{r} \right) dt^2 + \left( 1+\frac{r^2}{L^2} - \frac{2M}{r} \right)^{-1} dr^2.
\end{equation}
To perform the new thermodynamical analysis, the two-dimensional Euclidean Einstein-Hilbert action must be used:
\begin{equation} \label{eq:2d_EH}
S_{E}=-\int d^2 x \sqrt{-g}R.
\end{equation}
The Ricci scalar $R$ in the above action is defined using metric (\ref{eq:trunc}). The Euclidean action of dimensionally-reduced S-$\rm{AdS_{4}}$ spacetime can now be found from (\ref{eq:trunc}) and (\ref{eq:2d_EH}):
\begin{equation}
S_{\mathrm{EAdS-Sch}}=-\int_{0}^{\beta_{2}}dt \int_{r_{h}}^{\xi}dr R,
\end{equation}
giving
\begin{equation} \label{eq_comp1}
S_{\mathrm{EAdS-Sch}} = 2\beta_{2}\left( -\frac{r_{h}}{L^2} - \frac{M}{r_{h}^2} +\frac{M}{\xi^2} + \frac{\xi}{L^2} \right).
\end{equation}
The Euclidean action of the reduced $\rm{AdS_{4}}$ spacetime, described by metric (\ref{eq:trunc}) at $M=0$, is
\begin{equation} \label{eq_comp2}
S_{\mathrm{EAdS}} = -\int_{0}^{\beta_{1}}dt \int_{0}^{\xi}dr R = \frac{2\beta_{1}\xi}{L^2},
\end{equation}
and, as before, we must constrain the time periods using
\begin{equation}
\beta_{1}\sqrt{1+\frac{\xi^2}{L^2}}=\beta_{2}\sqrt{1-\frac{2M}{\xi}+\frac{\xi^2}{L^2}}
\end{equation}
in order to compare the free energies of the two spacetimes consistently.

The difference in actions at the asymptotic limit is using (\ref{eq_comp1}) and (\ref{eq_comp2})
\begin{equation}
\lim_{\xi \to \infty} \left( S_{\mathrm{EAdS-Sch}} - S_{\mathrm{EAdS}} \right) = -\frac{2\beta_{2}}{L^2 r_{h}^2}\left( r_{h}^3 + ML^2 \right).
\end{equation}
This is a new result showing that the S-$\rm{AdS_{4}}$ black hole in the dimensionally-reduced description is thermodynamically unfavoured only when the following holds: $r_{h}^3 < -ML^2$. As this can clearly never be satisfied, the black hole is completely stable, thermodynamically and hence topologically, in the dimensionally-reduced description. In this description, the Euclideanised spacetime has $\chi=1$ for any black hole temperature above its minimum value. This provides the reason why our two-dimensional formula (\ref{eq:2d_form}) gives the correct Hawking temperature after reduction whereas an attempt at a full four-dimensional description falters.

The effectiveness of temperature formula (\ref{eq:2d_form}) to other types of AdS black holes is demonstrated in the Appendix.

\section{Conclusions}

In this work we have shown that the topological method of determining Hawking temperatures is only applicable for S-$\rm{AdS_{4}}$ black holes after a dimensional reduction down to two dimensions. This is due to the Hawking-Page phase transition. This result, along with our previous work, suggests that the Hawking temperatures of a large number, and most probably all, four-dimensional black holes can be studied using a two-dimensional topological approach. This characteristic of black hole systems hints at a previously unfamiliar lower-dimensional encoding of horizon temperature information, reminiscent of the known area law of entropy and the holographic principle.

A thermodynamical analysis of the dimensionally-reduced S-$\rm{AdS_{4}}$ black hole spacetime was carried out, showing that the Hawking-Page phase transition is no longer present in this description. The phase transition in four dimensions destabilises the Euclideanised topology, making a manifold with vanishing Euler characteristic thermodynamically favoured. In the reduced description, the Euler characteristic is always non-zero, allowing a topological two-dimensional Hawking temperature formula to be employed.

It would be interesting to investigate further the topological description of black hole thermodynamics in future work, perhaps in more exotic and extreme spacetimes.

\section*{Acknowledgments}

L.D.M.V. acknowledges support from EPSRC (UK, Grant No. EP/L015110/1) under the auspices of the Scottish Centre for Doctoral Training in Condensed Matter Physics. F.B. acknowledges funding from the German Max Planck Society for the Advancement of Science (MPG), in the framework of the International Max Planck Partnership (IMPP) between Scottish Universities and MPG.

\section*{Appendix}

In the main text of this paper we have looked at only one type of $\rm{AdS_{4}}$ black hole, namely the S-$\rm{AdS_{4}}$ black hole with spherical event horizon. In fact, a whole family of different black holes exist in $\rm{AdS_{4}}$ spacetime with a wide variety of features. This family of black holes can be divided into three classes: those with spherical, hyperbolic, or flat event horizons. In this Appendix we will study the Hawking temperature using topology of two interesting anti-de Sitter black hole solutions: those with torus-shaped flat event horizons and those with negatively curved, hyperbolic horizons. One black hole from each class of the $\rm{AdS_{4}}$ family has therefore been studied using our topological method in this work, in each case giving correct temperature results.

The toral black hole in $\rm{AdS_{4}}$ spacetime has vanishing Euler characteristic due to its horizon's compactness and its genus value of $g=1$. A zero value for $\chi$ precludes a topological Hawking temperature study as discussed in the main text but its metric can be dimensionally reduced by simply removing the toral section, after which formula (\ref{eq:2d_form}) can be used.

The toral AdS black hole has a four-dimensional Euclidean metric \cite{Ong2}:
\begin{equation}
\begin{gathered}
ds^2 = \left( \frac{r^2}{L^2} - \frac{2M}{\pi K^2 r} \right) dt^2 + \left( \frac{r^2}{L^2} - \frac{2M}{\pi K^2 r} \right)^{-1} dr^2 \\ + r^2 \left( d\zeta^2 + d\xi^2 \right),
\end{gathered}
\end{equation}
where $K$ is a compactification parameter acting on the torus. As the torus has genus $g=1$, the Euler characteristic of the full manifold is zero however, crucially, dimensionally reducing to two dimensions gives
\begin{equation} \label{toral_reduc}
ds^2 = \left( \frac{r^2}{L^2} - \frac{2M}{\pi K^2 r} \right) dt^2 + \left( \frac{r^2}{L^2} - \frac{2M}{\pi K^2 r} \right)^{-1} dr^2,
\end{equation}
changing the topology to one with $\chi=1$.

Metric (\ref{toral_reduc}) can be processed by our temperature formula (\ref{eq:2d_form}) giving the correct result for the toral black hole \cite{Birm}
\begin{equation}
T_{\mathrm{H}}=\frac{3r_{h}}{4\pi L^2},
\end{equation}
where $r_{h}=\left( 2ML^2/\pi K^2 \right)^{1/3}$. 

Black holes are also known to exist in $\rm{AdS_{4}}$ spacetime having hyperbolic horizons. These seem rather unphysical (as their mass can reach negative values) however for completeness we treat them here.

We look at one particular example, treated in \cite{Ong2}, of an $\rm{AdS_{4}}$ black hole with compact hyperbolic horizon of genus $g=2$. This horizon form leads to a negative Euler characteristic: $\chi=2-2g=-2$. This spacetime is quite complicated, with the number of horizons depending on certain parameter values \cite{Vanzo}. Here, for simplicity, we choose a black hole with parameter values fixing the number of horizons to one and with a positive mass $M>0$.

The hyperbolic $\rm{AdS_{4}}$ black hole metric, after a dimensional reduction leaving only the $g_{rr}$ and $g_{tt}$ sections of the line element, is
\begin{equation} \label{eq:hyperb}
ds^2 = \left( -1+\frac{r^2}{L^2} - \frac{2M}{r} \right) dt^2 + \left( -1+\frac{r^2}{L^2} - \frac{2M}{r} \right)^{-1} dr^2.
\end{equation}
Dimensionally reducing from the full four-dimensional metric to (\ref{eq:hyperb}) alters the Euler characteristic from $\chi=-2$ to $\chi=1$. Again, a Ricci scalar $R$ can be found for this geometry, which when inputted into (\ref{eq:2d_form}) leads to the known Hawking temperature
\begin{equation}
T_{\mathrm{H}}=\frac{3r_{h}^2 - L^2}{4\pi L^2 r_{h}}.
\end{equation}

\end{document}